\documentclass[twocolumn]{aastex62} 
\usepackage[utf8]{inputenc}
\newcommand\degree{\degr} 

\received{\today}  
\accepted{...}     

\begin{document} 

   \title{Galaxy rotation curve from classical Cepheids} 
	 \author[0000-0003-0272-7791]{Piotr Gnaci\'nski} 
   \address{
          Institute of Theoretical Physics and Astrophysics,  
          Faculty of Mathematics, Physics and Informatics, \\
              University of Gda\'nsk, 
              ul. Wita Stwosza 57, \\ 
              80-308 Gda\'nsk, Poland 
    }

   \begin{abstract}  
      The Galaxy rotation curve is usually assumed to be flat.
	 However, some galaxies have rotation curves that are lower than the flat rotation curve.
	 In our Galaxy the Keplerian rotation of interstellar clouds 
	 in the galactic longitude $l=135\degree$ was observed.
	
	   We use a kinematic approach to derive the rotational velocity
	 of classical Cepheids. 
	 The rotational velocity was calculated from radial velocity and from proper motion.
	 The derived rotational velocities of Cepheids are between Keplerian 
	 rotation and the flat one.
	 We fit a Galaxy rotation model consisting of a black hole, bulge, disk and halo
	 to the rotation curve.
	 The density of dark matter halo is at least 60\% less than the value 
	 obtained from the flat rotation curve.
	
	 \end{abstract}  
 
	 \keywords{   
     Galaxy kinematics and dynamics -- dark matter
   }  
 
\section{Introduction}

The current consensus is a flat rotation curve of our Galaxy.
\cite{Reid} have obtained flat rotation curve basing on proper motions 
and trigonometric parallaxes of masers associated with high-mass star forming regions.
To sustain constant linear velocity dark matter must be present in our Galaxy. 

The term 'dark matter' first appeared
in the paper by \cite{Kepteyn}. Later \cite{Zwicky} has estimated
that the density of dark matter is grater than the amount of visible matter.
His estimations were based on velocities of galaxies in the Coma Cluster.
Today we observe dark matter due to gravitational lensing of galaxy clusters.
In cosmology, in the $\Lambda$CDM model we have 26\%  of density in dark matter, 
and only 5\%  of density in barionic matter \citep{Planck}.

	However, thare are many examples of rotation curves that are lower than the flat rotation curve.
\cite{Jalocha} have obtained mass distribution of the M 94 galaxy basing on
rotation curve, infrared luminosity and \ion{H}{1} 
observations.
The authors state that the obtained mass distribution {\it leaves no much room (if any) for dark matter}.
In a sample of 45 spiral galaxies analyzed by \cite{Honma} 11 galaxies have Keplerian rotation curve.
So about 1/4 of galaxies in their sample show Keplerian rotation.
 	
\cite{Sikora} have analyzed microlensing events in the inner part of our Galaxy.
The number of microlensing event is consistent with the amount of matter inferred
from Galaxy rotation curve. Authors wrote: {\it this result suggests
that non-barionic mass component may be negligible in this region}.

  \cite{Gazinur} have analyzed the Galactic rotation basing on interstellar clouds.
They have used interstellar \ion{Ca}{2}
absorption line to obtain both radial velocity and
distance. They have analyzed only the direction $l=135\degree$ and stated that the Galactic rotation
curve outside of the Solar orbit in that direction is Keplerian. 
	
	The Galactic rotation of old open clusters was analyzed by \cite{Gnacinski}. 
The rotation of old open clusters in the outer part of Galaxy agrees better with
Keplerian curve than with flat rotation curve. 
They proposed an explanation of various (flat/Keplerian) results of Galaxy rotation curves. 
If the orbits of objects are non-circular and the formula for rotation velocity 
is derived with the assumption of circularity then we get a very large spread of velocities, 
similar to the observed ones.
The non-circularity is justified by radial velocities measured in the Galactic anti-center.
Furthermore, a star located 23 kpc from Galaxy center needs 1 Gyr for 
one rotation around Galaxy (assuming Keplerian rotation).
So there may be to few revolutions to circularize the orbit.

  In this paper we analyze the rotational velocities of classical Cepheids
derived from radial velocities and from proper motions. 
We perform a fit of a Galaxy rotation model with dark
matter halo to the Cepheids rotational velocities.

\section{Data}

\begin{figure}
    \centering
		\includegraphics[width=\columnwidth]{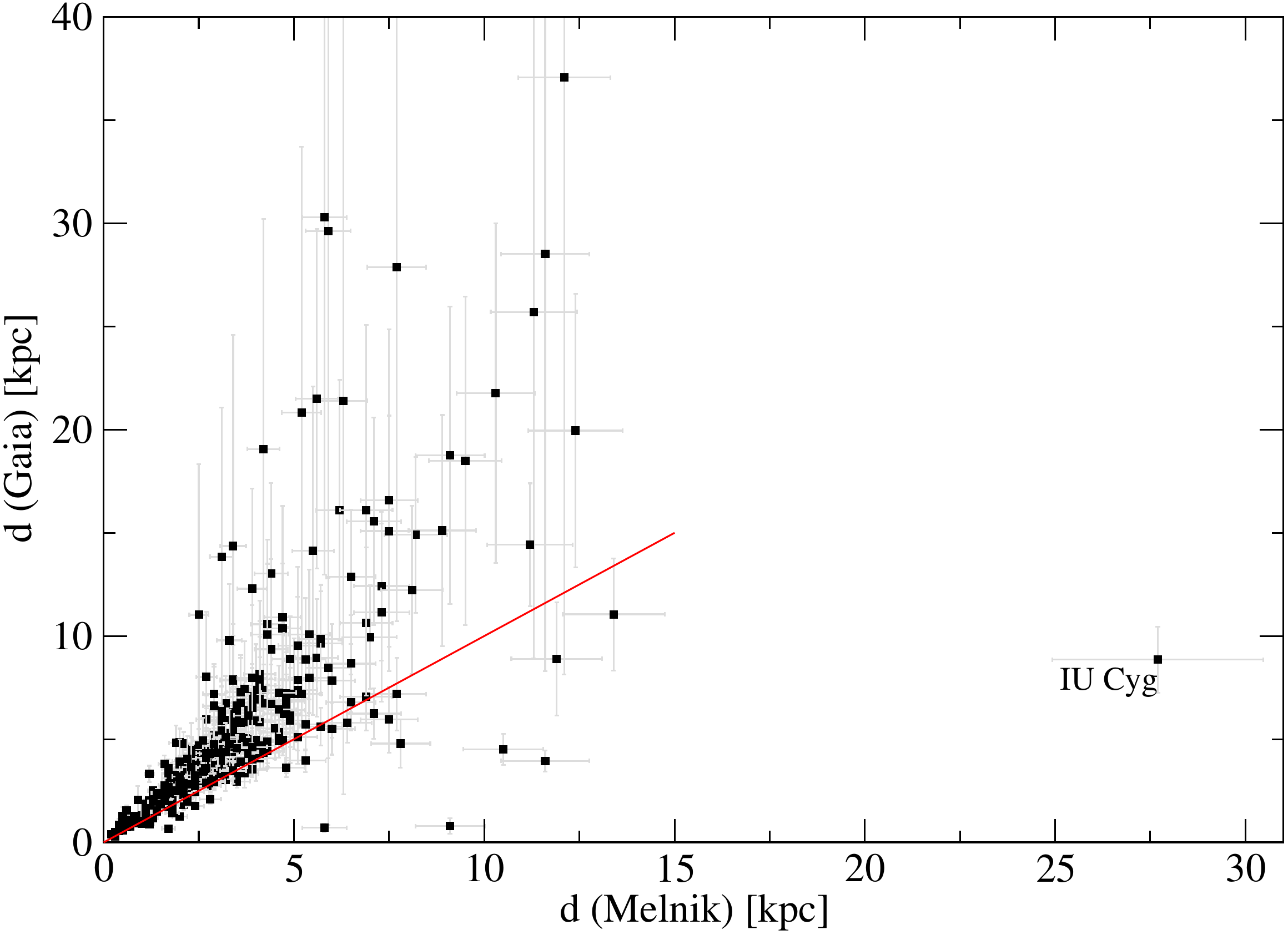}
    \caption{ Comparison of distances from \cite{Melnik} catalog and from Gaia DR2.
		   The line shows the equality relation.
		   We did not plot 18 problematic stars from the Gaia catalog:
			9 stars with negative parallaxes, 6 stars with Gaia distances greater than 40 kpc
			and 3 stars where the distance error was greater than the distance itself.
		}
    \label{fig_dist}
\end{figure}

This paper is based on Cepheids catalogue presented by \cite{Melnik}.
It is a compilation of measurements for 674 Cepheids.
The catalogue includes periods, V-bands magnitude, heliocentric distances,
heliocentric radial velocities and proper motions of classical Cepheids.

  The proper motion in the \cite{Melnik} catalog comes from the Hipparcos catalog.
We have used the proper motions from the latest Gaia Data Release 2 (DR2) catalog.
The distances to Cepheids from the \cite{Melnik} catalog were compared to the 
distances obtained from Gaia parallaxes (fig. \ref{fig_dist}).
There is a systematic difference between these distances.
The Gaia DR2 distances are usually higher than the distances from Cepheids catalog.
Moreover, there are problems with the Gaia distances, like negative parallaxes,
parallaxes with errors greater than the parallax itself or extremal large distances.
Therefore we have used distances from the \cite{Melnik} catalog.

  \cite{gaia} have analyzed the astrometric solution of Gaia DR2 data. 
They found a zero point offset of parallaxes -29~$\mu$as by analyzing the parallaxes of quasars.
The zero point offset depends on magnitude, color and position.
The comparison of Gaia DR2 parallaxes with external Cepheid catalogue was also performed by \cite{Arenou}.
The zero point offset of Gaia DR2 as compared to Cepheids distances equals to -31.9~$\mu$as.
They found also that there is no zero point offset of proper motion.

  The Gaia DR2 parallaxes of Cepheids were compared to parallaxes obtaind from HST photometry by \cite{Riess}.
The measured zero point offsets is -46~$\mu$as. 
The found also, that the Gaia DR2 parallaxes for bright Cepheids G$<$6~mag are unreliable because of Gaia's detectors saturation.

 We have used the Sun -- Galactic Center distance $R_{\Sun}=8\ \mathrm{kpc}$ and
the Sun velocity of $v_{\Sun} = 240\ \mathrm{km\,s}^{-1}$ 
(\citeauthor{Honma2012}, \citeyear{Honma2012}, \citeyear{Honma2015}; \citeauthor{Sofue2017}, \citeyear{Sofue2017}).
The rotation velocity of Cepheids was fitted with the \cite{Sofue2015} Galactic rotation
curve model.
The model consists of central black hole, exponential spherical bulge, exponential disk and halo 
(\citeauthor{Sofue2015}, \citeyear{Sofue2015}, \citeyear{Sofue2017}).
The halo density is given by
\begin{equation}
  \rho(R)= \frac{\rho_0}{\frac{R}{h}\left(1+{R}/{h}\right)^2}
\end{equation}
with the mass enclosed in radius R
\begin{equation}
  M_{halo}(R)= 4\pi\rho_0h^3\left(\ln(1+{R}/{h})-\frac{R/h}{1+R/h}\right).
\end{equation}

	We have used only the mass inside the Galactic center -- Sun distance to calculate the
Keplerian curve. 
Using $R_{\Sun}=8\ \mathrm{kpc}$ and $v_{\Sun} = 240\ \mathrm{km\,s}^{-1}$ we got
$M=1.07\cdot 10^{11}\ \mathrm{M_{\Sun}}$.
 The Keplerian curve is used for comparison on rotational velocity plots.

  According to \cite{Bovy} the masers associated with massive star forming regions that
	are located in the spiral arms are past the apocenter of their orbits.
	However, \cite{McMillan} argued that the systematic differences in maser velocities
	are caused by underestimation of solar velocity $V_\Sun$ in the LSR.
	We assume that no large--scale systematic motion is present in the Cepheids sample.
	Numerical simulations of Galaxy evolution by \cite{Baba} shows that star forming regions
	and young stars exhibit large non-circular motions.
	Their motion may be synchronized in a fragment of a spiral arm.
	Our stars are distributed uniformly in all directions and the average velocities should not
	be affected by small scale systematic motion.

\section{Rotation curve derived from radial velocity}

 \begin{figure}
    \centering
    \includegraphics[width=\columnwidth]{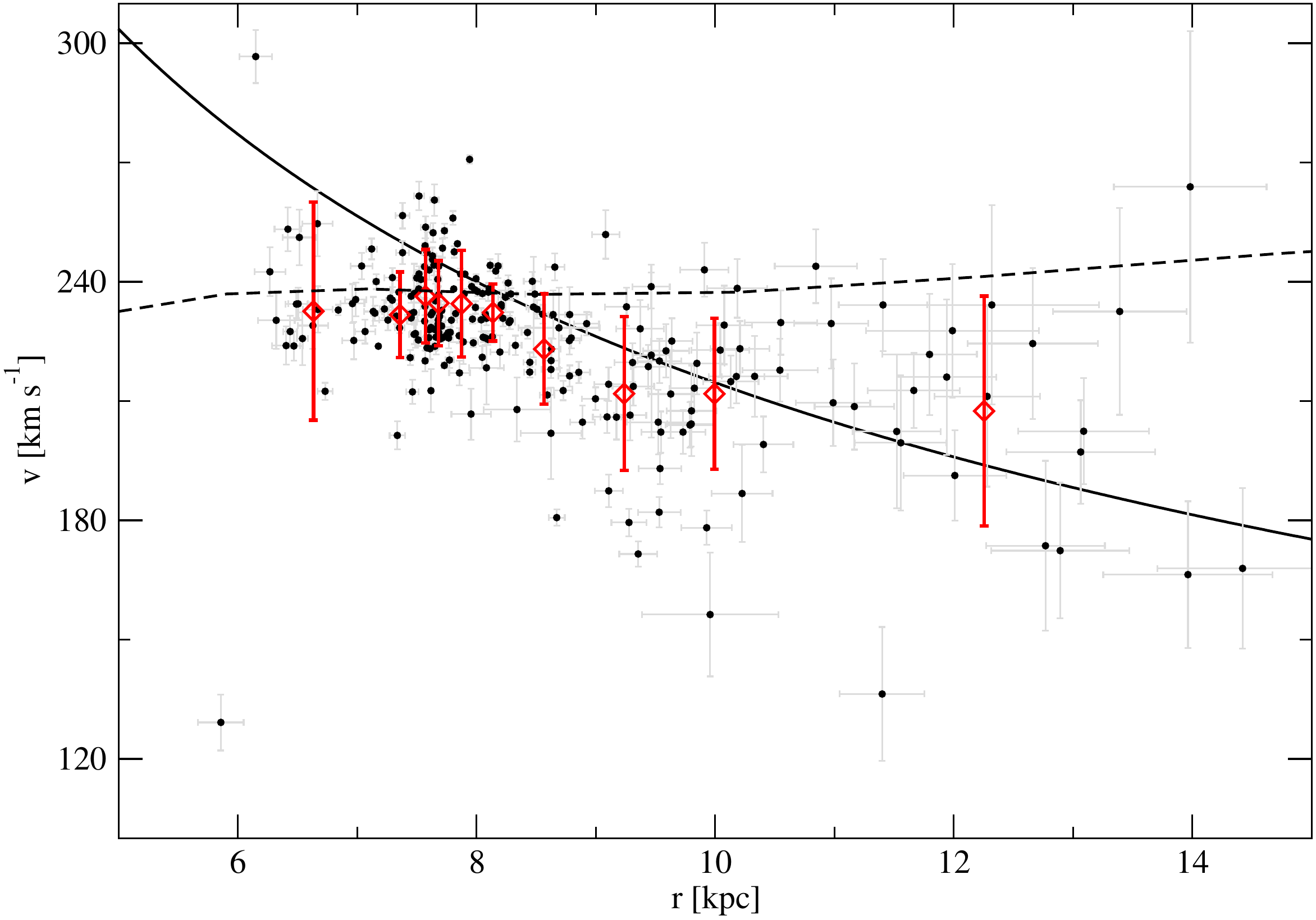}
    \caption{Linear rotation velocity of Cepheids derived from radial velocity.
						 Only stars with galactic longitude more distant than $30\degree$ from $180\degree$ and 
						 $0\degree$ are shown.
						 The vector of Solar motion is $(U,V,W)_{\Sun}$=(10,5.25,7.17).
						 Averages (diamonds) and their standard deviations are calculated from 24 successive points.
						 The solid curve represent Keplerian motion, while the dashed curve is the flat rotation curve
						 from \cite{Sofue2015}.
			}
    \label{rotVel_v}
 \end{figure}

 \begin{figure}
    \centering
    \includegraphics[width=\columnwidth]{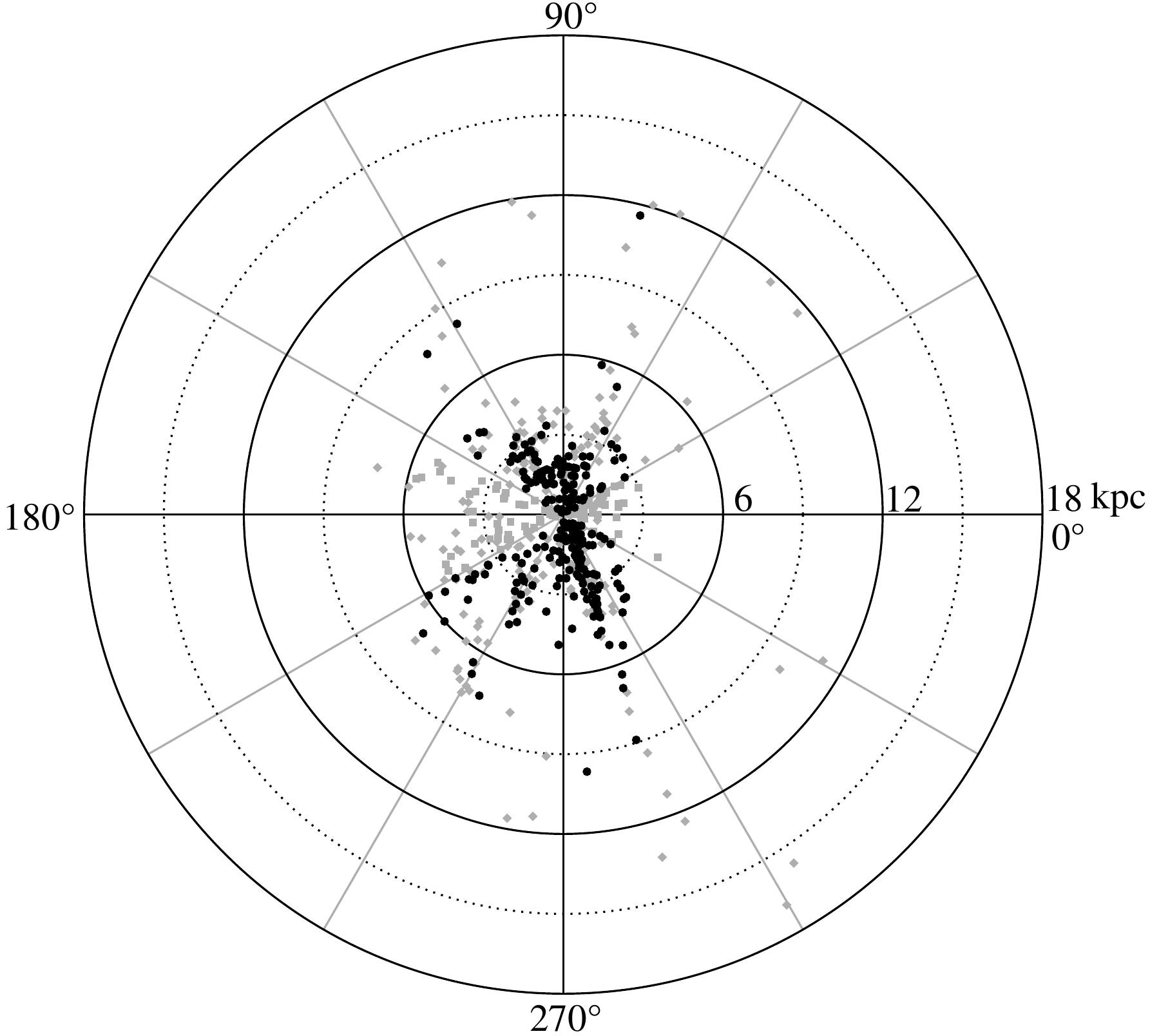}
    \caption{ Galactic longitudes and distances to Cepheids used to calculate the rotation velocity from radial velocity (black). 
		         The stars not used for rotation velocity calculation are shown in gray. They have galactic longitude 
						 closer than $30\degree$ to $180\degree$ or $0\degree$, or the systemic radial velocity
						 was not measured. The star IU Cyg ($l=69\fdg89$, $d=27.7$ kpc) is to distant to be shown on this plot.
						 This star does not have measurement of systemic radial velocity.
			}
    \label{position_Vrad}
 \end{figure}

Only 324 of 674 Cepheids catalogued by \cite{Melnik} have measurements of heliocentric radial velocities.
The line-of-sight velocities $v_h$ ($\gamma$-velocities) 
were converted to the Local Standard of Rest (LSR) velocities:
\begin{equation} 
     v_{LSR}=v_h + U_{\Sun} \cos b \cos l + V_{\Sun} \cos b \sin l + W_{\Sun} \sin b               \label{v_LSR}
\end{equation}
   where $(U_{\Sun},V_{\Sun},W_{\Sun})$ is the Sun velocity vector, and 
	 $(l,b)$ are Galactic coordinates of the star.
	 We have considered two values of Sun velocity in the Local Standard of Rest (LSR)
	 $(U,V,W)_{\Sun}=(11.1,12.24,7.25)\textrm{ km~s}^{-1}$ from
	 \cite{Schonrich}, and 
	 $(U,V,W)_{\Sun}=(10.00,5.25,7.17)\textrm{ km~s}^{-1}$ from
	 \cite{Dehnen}.

   The rotational velocity was calculated using formula derived for circular orbits \citep[eg.][]{Bhattacharjee}
   \begin{equation} 
     v=\frac{r}{R_{\Sun}} \left( \frac{v_\mathrm{LSR}}{\sin l \cos b} + v_{\Sun} \right).   \label{omega_r_z_vr}
   \end{equation}
   The star distance $r$ is the projection of galactocentric distance on the galactic plane
   \begin{equation} 
     r=\sqrt{R_{\Sun}^2 + d^2 \cos^2 b -2R_{\Sun} d \cos b \cos l},             \label{r}
   \end{equation}
	  where $d$ is the heliocentric distance to Cepheid.
		According to \cite{Melnik} the distances to Cepheids are known
		with the accuracy of about 10\%. We have used this value in our error calculations.

	Monte--Carlo simulations \citep[see][]{Gnacinski}
	of stars on elliptical orbits	with Keplerian motion shows that 
	even for huge eccentricities ($e$ distributed uniformly in the
	range $0\textrm{--}0.9$) the binned average of velocity stays 
	close to the assumed Keplerian rotation curve. This justifies
	that the calculation of average velocity makes sense even for
	non-circular motion.
	
The stars located near Galactic center ($l=0\degree$) and antycenter 
($l=180\degree$) must be excluded from the analysis, 
because the $\sin l$ in the denominator of eq. \ref{omega_r_z_vr} leads to
unphysical velocities (larger than the escape velocity from our Galaxy).
 No cutoff was performed basing on the distance to galactic plane $|z|$.
	
	The linear rotation velocities derived from radial velocities 
are presented on figure \ref{rotVel_v}. 
The rotation velocities were calculated using the 
$(U,V,W)_{\Sun}=(10,5.25,7.17)\textrm{ km~s}^{-1}$ from \cite{Dehnen}.
The values from \cite{Schonrich} $(U,V,W)_{\Sun}=(11.1,12.24,7.25)\textrm{ km~s}^{-1}$
gives velocities closer to the flat rotation curve.

The binned averages of velocity on figure \ref{rotVel_v} are calculated 
from 24 successive points. The averages are shown together with their standard deviations.
Note that stars with galactocentric distance less than $R_{\Sun}$ have average 
velocity close to flat rotation curve, 
while further stars have average velocities located close to the Keplerian rotation curve.
The distribution of analysed Cepheids in the Galactic disk is
shown on figure \ref{position_Vrad}.

\section{Rotation curve derived from proper motion}

\begin{figure}
    \centering
		\includegraphics[width=\columnwidth]{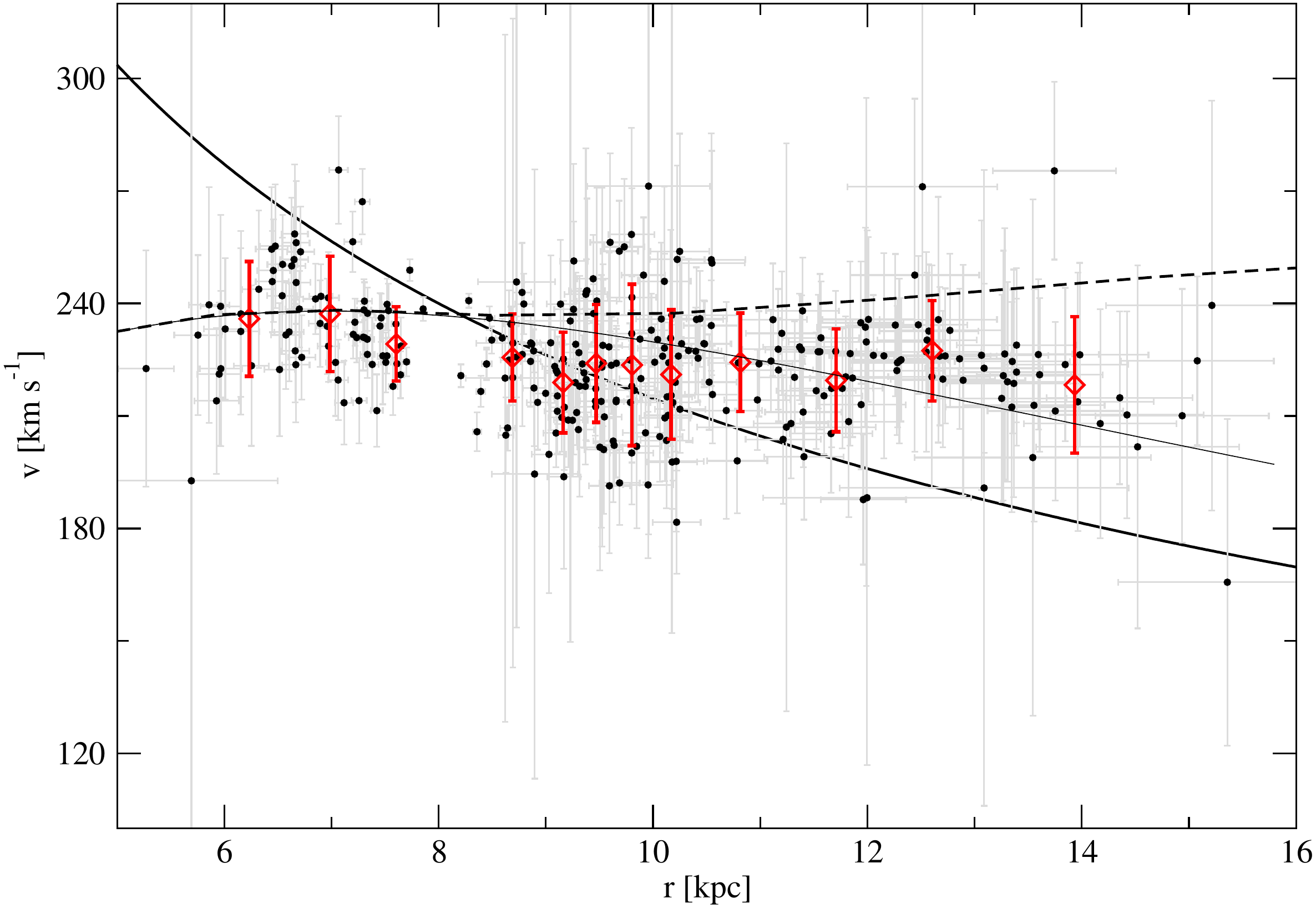}
    \caption{ Linear rotation velocity derived from proper motion. Only stars with $|R_{\Sun}\cos l -d|>4$ are shown.   
							Averages (diamonds) and their standard deviations are calculated from 25 successive points.
						  The solid curve represent Keplerian motion, while the dashed curve is the flat rotation curve
						  from \cite{Sofue2015}.
							The thin solid line is Galaxy rotation model with halo density equal to $\rho_0=0.00192$~M$_\Sun$\,pc$^{-3}$.
		}
    \label{fig_pm_v}
\end{figure}

 \begin{figure}
    \centering
    \includegraphics[width=\columnwidth]{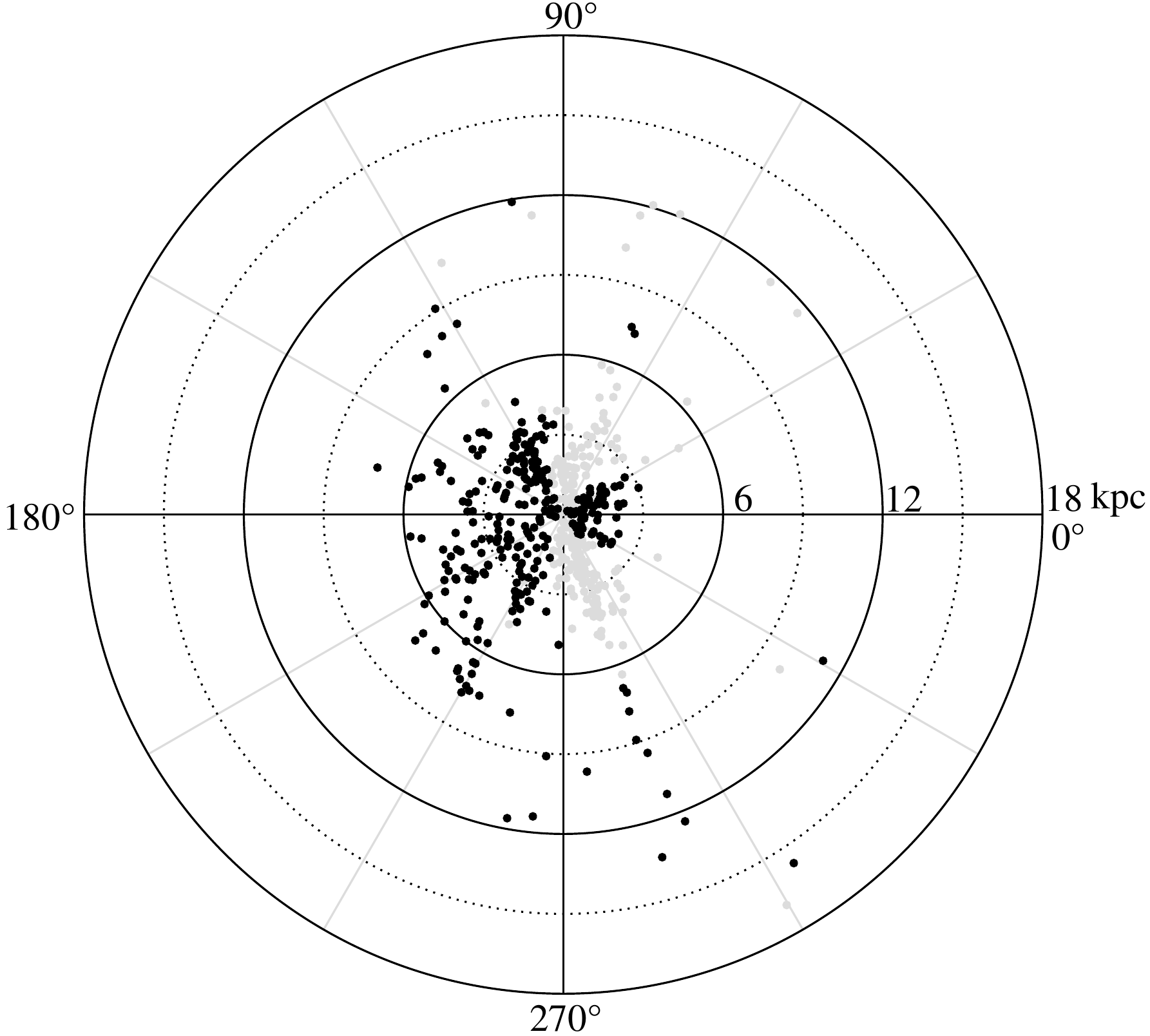}
    \caption{ Galactic longitudes and distances to Cepheids used to calculate the rotation velocity from proper motion (black). 
		         The stars not used for rotation velocity calculation are shown in gray. The outlier star IU Cyg is not shown on this plot.
			}
    \label{position-mi_l}
 \end{figure}

 The Gaia DR2 catalog \citep{gaia} has proper motions for 498
 Cepheids out of 674 in the \cite{Melnik} catalog.
 The velocity of Galaxy rotation was calculated using standard formula (derived
 with the assumption of circular orbits):
 
 \begin{equation} 
     v= r\frac{v_t+v_{\Sun} \cos l}{ R_{\Sun} \cos l - d },   \label{omega_r_mi}
  \end{equation}
	where $v_t=\mu_l^*\cdot d$ and $\mu_l^*$ is the proper motion in galactic longitude
multiplied by $\cos b$.
The stars with small denominator in eq. \ref{omega_r_mi} were excluded from the analysis, 
because small denominator leads to nonphysically large velocities.

Five stars were excluded from further analysis.
The star V979 Aql was excluded because of negative rotation velocity.
Stars IU Cyg, SS CMa, IX Cas and DQ And are outliers. Also 7 stars more
distant than 500 pc from the galactic plane were excluded.
We have used the Sun distance from Galactic plane $z_\Sun=17.4~pc$,
which is the median value from 56 estimations presented by \cite{Karim}.

  Figure \ref{fig_pm_v} presents linear 
rotation velocities obtained from Cepheids proper motion.
Averages and their standard deviation are calculated from 25 successive points.
Figure \ref{position-mi_l} shows the space distribution of Cepheids used
in our analysis. Note that the stars cover the region near $l=0\degree$ 
and $l=180\degree$ which was excluded in the radial velocity analysis (section 3).
The are exactly 100 stars common in the rotational velocity calculations 
from radial velocity and from the proper motion.

\section{3D velocity vector}

  For stars with radial velocities and proper motion we have calculated the three dimensional
	velocity vector using the formulas from \cite{Reid2009}. 
	There were 318 stars for which the velocity vector could be calculated.
	Four outliers were removed from further analysis: SS CMa, V Vel, AA Ser, V340 Ara.

\begin{figure}
    \centering
    \includegraphics[width=\columnwidth]{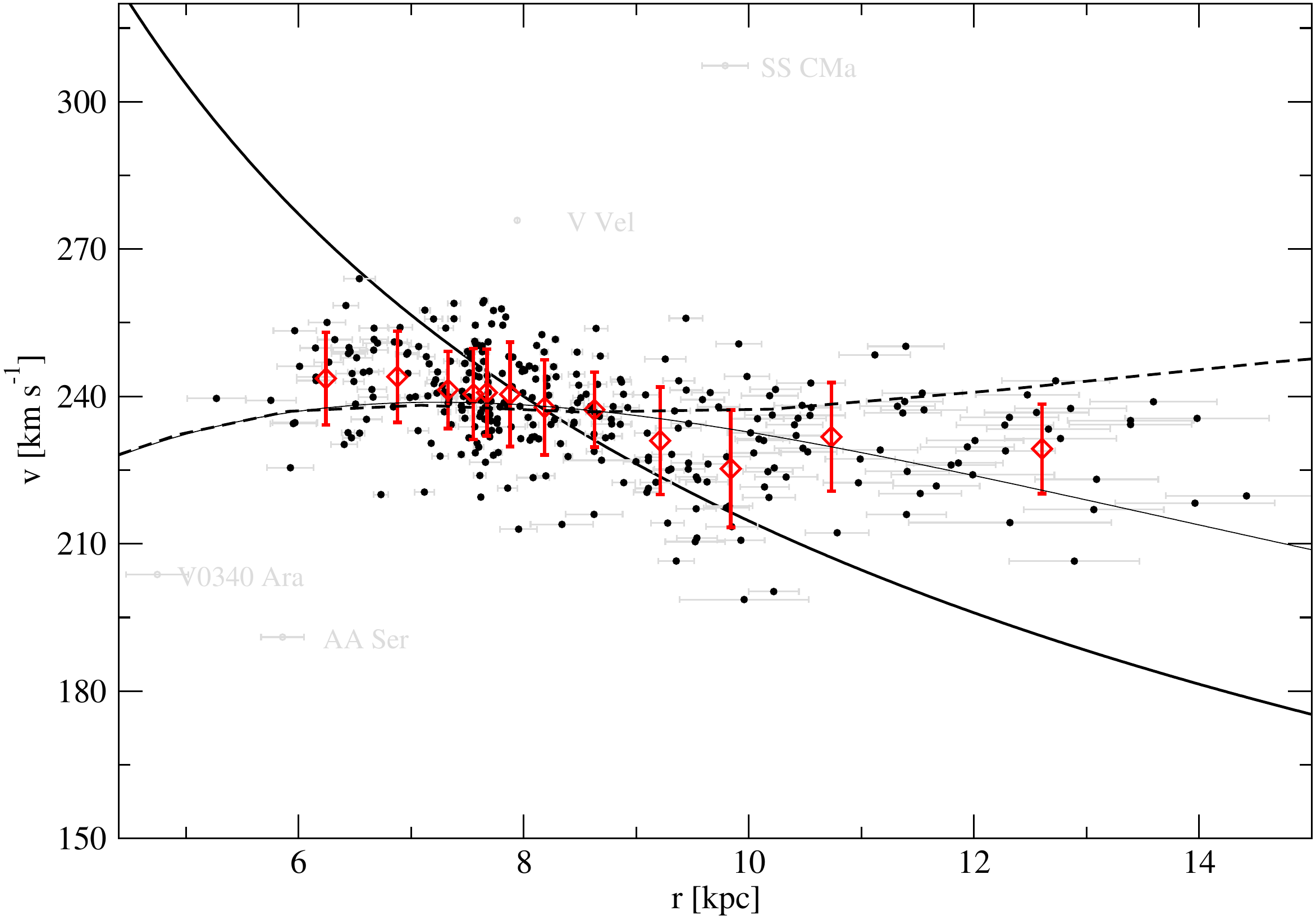}
    \caption{Circular rotation velocity of Cepheids from three dimensional velocity vector.
						 The vector of Solar motion is $(U,V,W)_{\Sun}$=(11.1,12.24,7.25).
						 Averages are calculated from successive 26 points.
						 The solid curve represent Keplerian motion, while the dashed curve is the flat rotation curve
						 from \cite{Sofue2015}.
						 The thin solid line is Galaxy rotation model with halo density equal to $\rho_0=0.0063~M_\Sun\,pc^{-3}$.
			}
    \label{circ_vel}
 \end{figure}

\begin{figure}
    \centering
		\includegraphics[width=\columnwidth]{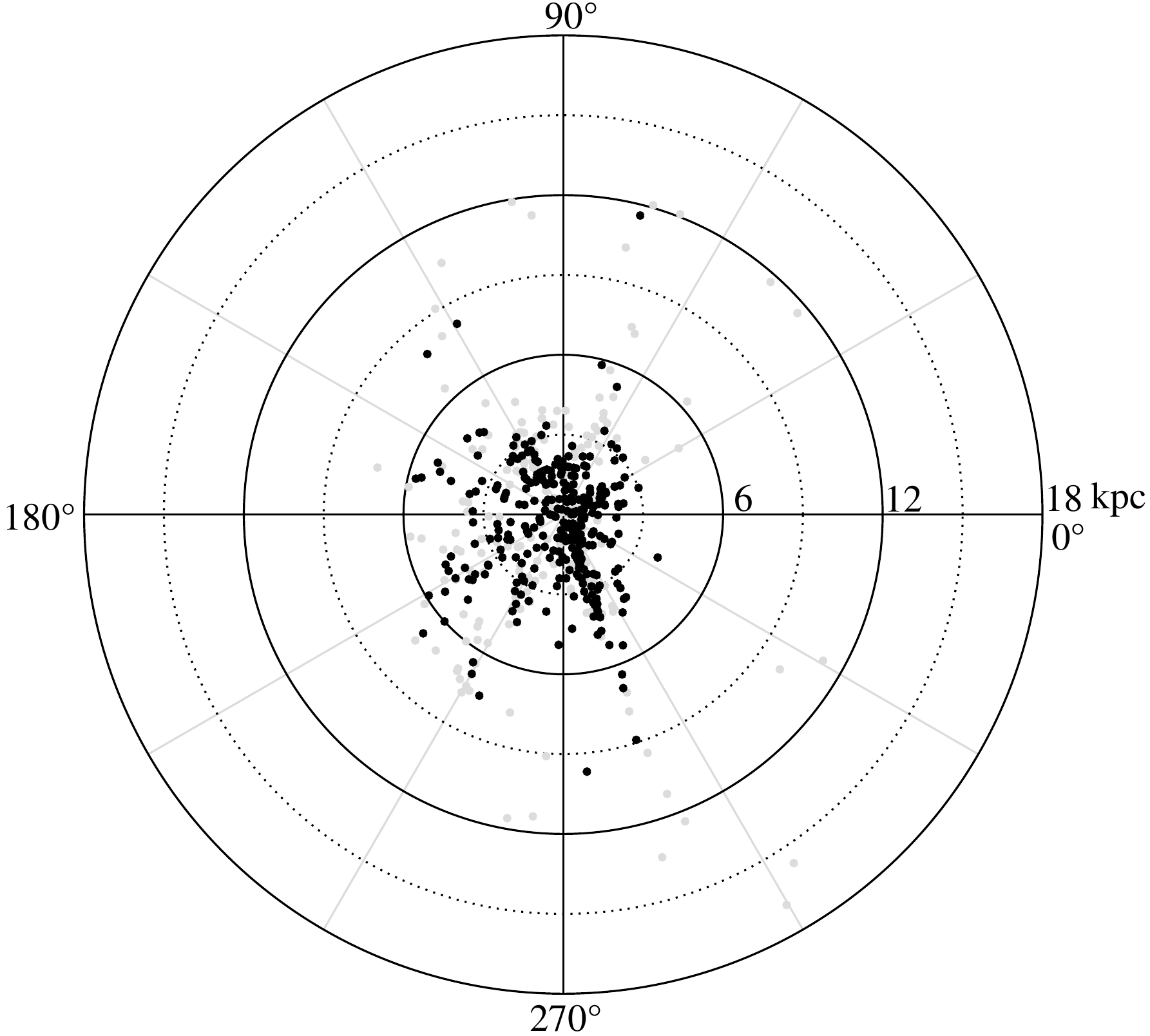}
    \caption{ Galactic longitudes and distances to Cepheids used to calculate the circular velocity (black). 
		         The stars without radial velocity or proper motion are shown in gray. The outlier star IU Cyg is not shown on this plot.
		}
    \label{fig_poz_6D}
\end{figure}

  The 3D rotation velocity vector was calculated once using the
	Sun velocity in the Local Standard of Rest (LSR)
	 $(U,V,W)_{\Sun}=(11.1,12.24,7.25)\textrm{ km~s}^{-1}$ from
	 \cite{Schonrich}, and second time using
	 $(U,V,W)_{\Sun}=(10.00,5.25,7.17)\textrm{ km~s}^{-1}$ from
	 \cite{Dehnen}. 
	The circular velocities of 3D velocity vector are presented on figure \ref{circ_vel}.
	Averages were calculated for 26 successive points.
	The positions of Cepheids with 3D velocity vector are shown on figure \ref{fig_poz_6D}.

\section{Results and discussion}

\begin{table*}
 \centering
 \begin{tabular}{lrrrr}   
 \hline         
 Data & stars &	\multicolumn{1}{c}{$\rho_0~[M_\Sun\,pc^{-3}]$} & $\rho_0/\rho_0^{flat}$ \\ 
 \hline
  V$_{circ}$ from 3D V vector $(U,V,W)_{\Sun}=(10.0,5.25,7.17)\textrm{ km~s}^{-1}$ & 160	& $<1.5\cdot 10^{-3}$ & $<$8\% \\
  V$_{circ}$ from 3D V vector $(U,V,W)_{\Sun}=(11.1,12.24,7.25)\textrm{ km~s}^{-1}$  & 160 & $7.3\pm1.3\cdot 10^{-3}$ & 40\% \\
	V from proper motion &  &  \\
	$|R_{\Sun}\cos l - d|>4$ AND $z<500~pc$ & 228 & $2.8\pm1.0\cdot 10^{-3}$ & 15\% \\
  V from radial velocity $(U,V,W)_{\Sun}=(10.0,5.25,7.17)\textrm{ km~s}^{-1}$ &  \\
	$|l-180\degree|>30\degree$ AND $l>30\degree$ AND $l<330\degree$ & 120 & $<4\cdot 10^{-6}$ & -- \\
  V from radial velocity $(U,V,W)_{\Sun}=(11.1,12.24,7.25)\textrm{ km~s}^{-1}$  & 	\\
	$|l-180\degree|>30\degree$ AND $l>30\degree$ AND $l<330\degree$ & 120 & $<9\cdot 10^{-6}$ & -- \\
 \hline
 \end{tabular}
 \caption{ Halo density derived from fitting Galaxy rotation model to the rotational velocities. 
           Only velocities from stars with $R>R_{\Sun}$ were fitted. 
					 The $\rho_0^{flat}$ from \cite{Sofue2015} equals to 0.0182~$M_\Sun\,pc^{-3}$.
					}
 \label{tab_ro0}
\end{table*}

Radial velocity, proper motions and circular component of the 3D velocity vector leads to
rotation velocities that are between the Keplerian rotation curve 
and the flat one. 
The outer parts of our Galaxy rotate slower than the flat rotation curve.

Monte-Carlo simulations show, that the rotation velocity derived from 
proper motion in the case of non-circular motion is more accurate, than
the rotation velocity derived from radial velocity. 
Although the 3D velocity gives us the whole information about the velocity vector, 
the rotational velocity derived from proper motion gives us larger sample of stars.
Some of the additional stars are located further from galactic center than the stars
for which we have full 3D velocity information.

We have constructed Galaxy rotation data consisting of \cite{Sofue2015} rotation curve up to
8~kpc from the galactic center and from individual velocities for $R>8$~kpc. 
The velocities were calculated from radial velocities, proper motion and 3D velocity vector. 
We use the inner rotation curve from \cite{Sofue2015}, because the tangent point method
used to determine rotation velocity is the most accurate one. 
We give the weight 1000 to the points from \cite{Sofue2015} rotation curve,
because we want to match the Sofue's model in the region $R<8$~kpc.

The $\rho_0$ parameter cannot be estimated from the inner rotation curve, because in
this region the rotational velocity is dominated by stellar components (bulge and disk).
We have kept the halo scale radius fixed to $h=10.7$~kpc as given by \cite{Sofue2015}. 
 
  Various galaxy components influence the rotation velocity at different distances from
	the Galaxy center.
	Therefore, the fitting of rotation curve was performed in three steps. 
	First we fitted the central black hole and bulge. 
	Next the bulge and disk up to 8~kpc.
	Then the disk from 2.7~kpc and our velocities were fitted.
	This procedure was repeated with different distance boundaries to fit the bulge, 
	leading to different bulge masses.
	The resulting halo densities are presented in table \ref{tab_ro0}.
	
	In the case of circular velocity derived from 3D velocity vector and solar motion from \cite{Dehnen}
	we got halo density variations of 4 orders of magnitude.
	Therefore we give only the upper limit of $\rho_0$ in this case.
	
	The rotational velocities derived from radial velocities lead to rotation curves
	with almost zero halo density. However this curves are very sensitive to
	noncircular motion.
	
	The rotational velocities from proper motions are less sensitive to noncircular motion \citep{Gnacinski}.
	It is also the biggest sample in our analysis -- 228 stars further than $R_\Sun$.
	The Galaxy rotation model fitted to proper motion velocities has halo density $\rho_0=(2.8\pm1.0)\cdot 10^{-3}~M_\Sun\,pc^{-3}$, 
	that is	15\% of the \cite{Sofue2015} value.
	
	Also the fittings of Galaxy rotation model to circular component of 3D velocity vector  
	(solar LSR velocity from \citealp{Schonrich}) gives repeatable
	results with $\rho_0=(7.3\pm1.3)\cdot 10^{-3}$~M$_\Sun$\,pc$^{-3}$. 
	It is 40\% of halo density by \cite{Sofue2015} for the flat rotation curve.
	At the Sun distance this halo density equals to $\rho(R_\Sun)=3.2\cdot 10^{-3}$~M$_\Sun$\,pc$^{-3}$ or $0.12$~GeV\,cm$^{-3}$.

  The inner rotation curve is usually determined using the tangent point method.
	Numerical simulations by \cite{Chemin} shows that the bump at the rotation curve
	below 1~kpc may be caused by the Galaxy bar and does not reflect the true rotation velocity.
	However, for distances larger than 5~kpc from Galaxy center both the rotation curve and
	velocity from bulge are not affected by the bar placement.
	Since we fit our velocity points for $R>8$~kpc the details of bulge/bar placement are 
	not significant for our results.

 The analysis of Hipparcos proper motions of 220 Cepheids by \cite{Feast} suggests that the
Galactic rotation curve declines at the solar distance. 
The dynamical approach to derive rotation curve was used by \cite{Eilers}. 
They used over 23~000 red giants with APOGEE, WISE, 2MASS and Gaia data.
Their velocity curve is declining and the derived local dark matter density equals to $\rho(R_\Sun)=0.3$~GeV\,cm$^{-3}$.

Recently \cite{Mroz} have also analyzed Galaxy rotation curve basing on classical Cepheids. 
Their rotation curve is also located below the Sofue's rotation curve, and
the outer rotation curve is declining. They did not estimate the halo density.
They use median Gaia velocities instead of systemic radial velocities of Cepheids.
The difference between the median Gaia velocities used by \cite{Mroz} and systemic
velocities given by \cite{Melnik} varies between -25 and +46.7~km/s. 

\section{Conclusions}
  
	Our kinematic approach to Galaxy rotation curve leads to declining rotation curve, that
	is located below the flat rotation curve by \cite{Sofue2015}.
	
 The main results are:
\begin{itemize}
	\item The rotation velocities of classical Cepheids derived from radial
	      velocities, from proper motion (Gaia DR2) and from three dimensional velocity vector
				are located between the flat and Keplerian rotation curves.
	\item The Galaxy rotation model fitted to the rotation velocities 
	      gives halo density at least 60\% less than the density
				calculated from the flat rotation curve.
	\item Lower values of $V_\Sun$ in the LSR leads to lower (or zero) halo density.
  \end{itemize} 
 
\subsection*{Acknowledgments}
 This research has made use of the SIMBAD database, operated at CDS, Strasbourg, France \citep{Wanger}.
 This work has made use of data from the European Space Agency (ESA) mission
{\it Gaia} (\url{https://www.cosmos.esa.int/gaia}), processed by the {\it Gaia}
Data Processing and Analysis Consortium (DPAC,
\url{https://www.cosmos.esa.int/web/gaia/dpac/consortium}). Funding for the DPAC
has been provided by national institutions, in particular the institutions
participating in the {\it Gaia} Multilateral Agreement.

\end{document}